\documentclass[pdflatex,sn-mathphys-num]{sn-jnl}

\usepackage{graphicx}%
\usepackage{multirow}%
\usepackage{amsmath,amssymb,amsfonts}%
\usepackage{amsthm}%
\usepackage{mathrsfs}%
\usepackage[title]{appendix}%
\usepackage{xcolor}%
\usepackage{textcomp}%
\usepackage{manyfoot}%
\usepackage{booktabs}%
\usepackage{algorithm}%
\usepackage{algorithmicx}%
\usepackage{algpseudocode}%
\usepackage{listings}%
\usepackage[normalem]{ulem}

\usepackage[T1]{fontenc}%
\usepackage[english]{babel}%
\usepackage{bm}%
\usepackage{braket}%

\begin{document}

\title[Portfolio Optimization via FALQON Variants]{Portfolio Optimization via FALQON Variants: Warm-Starting, XY Mixing, and Circuit-Depth Reduction}

\author*[1]{\fnm{Pedro M.} \sur{Prado}}\email{pm.prado@unesp.br}

\author[1]{\fnm{Rafael Sim\~oes do} \sur{Carmo}}

\author[2]{\fnm{Lucas A. M.} \sur{Rattighieri}}

\author[3]{\fnm{Luiz Gustavo Esmenard} \sur{Arruda}}

\author[1]{\fnm{Giovanni S.} \sur{Franco}}

\author[2]{\fnm{Marcos C\'esar de} \sur{Oliveira}}

\author[1,3]{\fnm{Felipe F.} \sur{Fanchini}}

\affil[1]{\orgdiv{School of Sciences}, \orgname{S\~ao Paulo State University (UNESP)}, \orgaddress{\city{Bauru}, \postcode{17033-360}, \state{SP}, \country{Brazil}}}

\affil[2]{\orgdiv{Instituto de F\'isica Gleb Wataghin}, \orgname{Universidade de Campinas}, \orgaddress{\city{Campinas}, \postcode{13083-970}, \state{SP}, \country{Brazil}}}

\affil[3]{\orgname{QuaTI -- Quantum Technology \& Information}, \orgaddress{\city{S\~ao Carlos}, \postcode{13560-161}, \state{SP}, \country{Brazil}}}

\abstract{
Discrete portfolio selection with cardinality constraints can be formulated as a quadratic unconstrained binary optimization (QUBO) problem. We investigate FALQON, TR-FALQON, and SO-FALQON combined with transverse-field, XY, and warm-start mixers. Statevector simulations were performed for $n=6$ assets over 25 random seeds and for a representative $n=10$ instance. \textcolor{black}{For $n=6$, the XY mixer achieved success probabilities above $97\%$, whereas for $n=10$, the warm-start mixer was the only approach to exceed the $50\%$ threshold. These results show that mixer design plays a central role in convergence and solution visibility.}
}


\maketitle

\section{Introduction}\label{sec:introduction}

Modern Portfolio Theory (MPT), as formalized in \cite{Markowitz1952}, frames asset allocation as a mean--variance trade-off: one seeks portfolios that minimize risk for a target return (or equivalently maximize return for a target risk). In large-scale institutional applications, this formulation is typically extended with practical constraints such as budgets, cardinality limits, concentration bounds, and transaction costs. Introducing binary decision variables for inclusion yields a non-convex quadratic optimization problem that can be written as a \emph{Quadratic Unconstrained Binary Optimization} (QUBO) instance \cite{Glover2018_QUBO_Tutorial, Kochenberger2014_UBQP_Survey}. Since QUBO is NP-hard \cite{Bienstock1996, Kochenberger2014_UBQP_Survey}, instances with only a few dozen assets can already challenge exact branch-and-bound approaches and higher-order convex relaxations.

In this context, variational quantum algorithms on \textcolor{black}{Noisy Intermediate-Scale Quantum (NISQ)} devices \textcolor{black}{\cite{preskill2018quantum}} are being investigated as potential heuristics for combinatorial optimization. The Quantum Approximate Optimization Algorithm (QAOA) \cite{Farhi2014} is a widely used baseline, but its quantum--classical outer loop introduces latency and may face difficult parameter landscapes in practice. The \emph{Feedback-Based Algorithm for Quantum Optimization} (FALQON) \cite{Magann2022} removes the classical parameter-search loop by employing a Lyapunov-inspired feedback rule that updates layer parameters directly from expectation values measured during execution. Subsequently, a second-order extension (SO-FALQON) was proposed, incorporating a quadratic approximation in the feedback rule to improve stability and convergence in regimes where first-order updates may be suboptimal \cite{Arai2025}. More recently, it was shown that time-rescaling the dynamics (TR-FALQON) can reduce the circuit depth required for convergence on QUBO problems, including \textsc{MaxCut} \cite{Rattighieri2025}, which is particularly relevant under limited coherence; related improvements from FALQON variants have also been reported in other QUBO mappings such as de novo DNA assembly \cite{Prado2026_DNAFALQON}.


Algorithmic performance is also strongly influenced by the initial state and by the choice of mixer Hamiltonian. Warm-starting \cite{Egger2021} prepares a coherent initial state using approximate solutions from continuous relaxations (QP or SDP), biasing the evolution toward promising regions of the landscape. Separately, for budget-constrained problems it is often beneficial to employ mixers that preserve Hamming weight; the XY mixer \cite{Wang2020XY, carmo2025warm} conserves the total number of excitations and can yield a more structured search within the feasible subspace than the conventional transverse $X$ mixer \cite{Hadfield2017_QAOAplus, Fuchs2022_ConstrainedMixers}. Complementary to warm-starting, measurement-guided strategies can iteratively refine a biased product initialization using statistics from previous shallow runs, improving feedback-based optimization performance without reintroducing a classical parameter-search loop \cite{Rattighieri2026_MGI}. In this work, we study these ingredients \emph{as distinct design choices} and benchmark three mixer constructions (transverse $X$, XY, and a warm-start-driven mixer) across three feedback-based optimizers (FALQON, TR-FALQON, and SO-FALQON), totaling nine configurations.

Beyond asset allocation, quantum routines have also been proposed for risk analysis, notably for estimating Value at Risk (VaR) and Conditional Value at Risk (CVaR) via amplitude amplification \cite{WoernerEgger2019}. Together, such primitives outline a broader quantum finance pipeline, from risk estimation to discrete portfolio selection \cite{orus2019quantum, egger2020quantum, herman2022survey, Kerenidis2019_Portfolio, Sakuler2023_RealWorldQA_Portfolio}.

This article is organized as follows. In Sec.~\ref{sec:portfolio} we cast the portfolio problem in QUBO form. In Sec.~\ref{sec:falqon} we present FALQON and specify the mixer constructions used in our comparisons. Sec.~\ref{sec:trso} presents the variants of FALQON. Sec.~\ref{sec:implementation} describes the experimental setup. Numerical results are presented in Sec.~\ref{sec:results}, and Sec.~\ref{sec:conclusion} closes with conclusions and perspectives for future work.

\section{Discrete Portfolio Problem Formulation}\label{sec:portfolio}

Consider a universe of $n$ assets described by the vector of mean returns $\boldsymbol{\mu}\in\mathbb{R}^{n}$ and covariance matrix $\boldsymbol{\Sigma}\in\mathbb{R}^{n\times n}$, estimated from historical data or factor models. The objective is to construct a long-only portfolio composed of exactly $K$ assets ($1\ll K\ll n$), all equally weighted with weight $1/K$. For each asset $i$ we introduce a binary decision variable
\[
x_i=
  \begin{cases}
  1,&\text{if asset $i$ is included},\\
  0,&\text{otherwise},
  \end{cases}
\]
so that $\boldsymbol{x}=(x_1,\dots,x_n)^{\mathsf T}$ specifies the selected portfolio.

In the mean--variance formulation of Markowitz \cite{Markowitz1952}, the portfolio risk is $ \boldsymbol{w}^{\mathsf T}\boldsymbol{\Sigma}\boldsymbol{w}$, while the expected return is $ \boldsymbol{w}^{\mathsf T}\boldsymbol{\mu}$. Under equal weights $w_i=x_i/K$, we apply the classical quadratic scalarization with risk-aversion coefficient $q>0$, obtaining the integer program
\begin{align}
\min_{\boldsymbol{x}\in\{0,1\}^{n}}
        &\; q\frac{1}{K^{2}}\boldsymbol{x}^{\mathsf T}\boldsymbol{\Sigma}\boldsymbol{x}
           -\frac{1}{K}\,\boldsymbol{\mu}^{\mathsf T}\boldsymbol{x},  \label{eq:mv_obj}\\[2pt]
\text{subject to}\;
        &\; \sum_{i=1}^{n} x_i = K.                           \label{eq:mv_card}
\end{align}

The cardinality constraint \eqref{eq:mv_card}, combined with the binary variables, makes the problem NP-hard \cite{Bienstock1996,Kochenberger2014_UBQP_Survey}. To make it suitable for quantum algorithms, we incorporate \eqref{eq:mv_card} into the objective via a quadratic penalty with coefficient $\lambda>0$ \cite{Glover2018_QUBO_Tutorial}, yielding the unconstrained cost

\begin{equation}\label{eq:qubo_port}
H_{\textsc{qubo}}(\boldsymbol{x})=
     q\frac{1}{K^{2}}\boldsymbol{x}^{\mathsf T}\boldsymbol{\Sigma}\boldsymbol{x}
   -\frac{1}{K}\,\boldsymbol{\mu}^{\mathsf T}\boldsymbol{x}
   +\lambda\Bigl(\sum_{i=1}^{n}x_i-K\Bigr)^{2}.
\end{equation}

Choosing $\lambda$ sufficiently large ensures that any violation of the cardinality constraint is energetically penalized \cite{Glover2018_QUBO_Tutorial}. The function \eqref{eq:qubo_port} is a quadratic polynomial in $0$--$1$ variables, i.e., a QUBO instance \cite{Glover2018_QUBO_Tutorial,Kochenberger2014_UBQP_Survey}. Applying the standard substitution $x_i \mapsto \tfrac{1}{2}(1-Z_i)$ yields an Ising Hamiltonian \cite{Glover2018_QUBO_Tutorial,Kochenberger2014_UBQP_Survey}

\begingroup
\color{black}
\begin{equation}
\hat{H}= \sum_{i<j} J_{ij}\,Z_iZ_j \;+\; \sum_{i} h_i\,Z_i \;+\; \text{const},
\end{equation}
\endgroup

where the couplings $J_{ij}$ originate from the quadratic (risk and penalty) terms and the fields $h_i$ from the linear return terms.

This representation is directly compatible with quantum algorithms \cite{Hadfield2017_QAOAplus}: $\hat{H}$ defines the \emph{problem Hamiltonian}, while a \emph{mixer} (driver) Hamiltonian generates exploration. A warm-start initial state can be prepared from inclusion probabilities obtained via continuous relaxations (QP or SDP), as detailed in \cite{Egger2021}. Additionally, the XY mixer preserves Hamming weight and therefore maintains the cardinality constraint provided the evolution is initialized within the $K$-excitation subspace; it can also induce optimization landscapes more benign than the conventional $X$ mixer \cite{Wang2020XY,Hadfield2017_QAOAplus,Fuchs2022_ConstrainedMixers}. This construction defines the discrete portfolio instances used as testbeds for the FALQON, TR-FALQON, and SO-FALQON variants studied in this work.

\section{FALQON}\label{sec:falqon}

The \emph{Feedback-Based Algorithm for Quantum Optimization} ~\cite{Magann2022} is a quantum algorithm designed to solve combinatorial optimization problems by steering the quantum state evolution through real-time feedback. Unlike traditional variational algorithms such as VQE and QAOA, which rely on iterative classical parameter optimization, FALQON offers a fully deterministic, gradient-free alternative that adapts the evolution path based on quantum measurements performed during execution.

Originally inspired by discrete-time \emph{Quantum Lyapunov Control}~\cite{Grivopoulos2004}, FALQON replaces the “measure–optimize–recompile” loop of QAOA with an \emph{in situ} feedback mechanism derived from stability criteria. FALQON operates by applying a time-dependent Hamiltonian that interpolates between an initial (easy-to-prepare) Hamiltonian $\hat H_d $ and a final (problem-specific) cost Hamiltonian $\hat H_p $. The quantum state $ |\psi(t)\rangle $ evolves according to the Schrödinger equation \textcolor{blue}{($\hbar=1$)}:

\begin{equation}
\frac{d}{dt}|\psi(t)\rangle = -i \left[ \hat H_p + \beta(t) \hat H_d \right] |\psi(t)\rangle,
\end{equation}
where the function $ \beta(t) $ is dynamically adjusted based on measurements of the current quantum state. Specifically, at each discrete time step, FALQON adjusts $ \beta(t) $ using a classical feedback rule that promotes convergence toward low-energy configurations. To be more specific, at each discrete time step $ t $, FALQON applies two short Trotterized evolutions of duration $ \Delta t $:

\begin{enumerate}
  \item Deterministic propagation under the cost Hamiltonian:  

        \begingroup
        \color{black}
        \begin{equation}
        \hat U_p(\Delta t) = \exp\bigl(-i\,\Delta t\,\hat H_p\bigr);
        \end{equation}
        \endgroup
        
  \item Controlled propagation under the mixer Hamiltonian:  
        \begingroup
        \color{black}
        \begin{equation}
        \hat U_d(\beta_t \Delta t) = \exp\bigl(-i\,\beta_t \, \Delta t\,\hat H_d\bigr).
        \end{equation}
        \endgroup
        \end{enumerate}

After applying the $ t $-th layer, the algorithm measures the functional

\begingroup
\color{black}
\begin{equation}
A_t \;=\;
\bigl\langle \psi_t \bigm| \,i[\hat H_d,\hat H_p] \,\bigm| \psi_t \bigr\rangle,
\end{equation}
\endgroup

which quantifies the instantaneous rate of change of the cost energy along the $\hat H_d$ direction of the dynamics. The next rotation angle is determined via a feedback rule:

\begin{equation}
    \beta_{t+1} \;=\; -\,A_t,
    \label{eq:lei de feedback1}
\end{equation}
ensuring that the time derivative of the expected cost energy remains non-positive and thus guarantees monotonic convergence to lower-energy states~\cite{Larsen2024}.

This feedback loop eliminates the need for classical search over a $2p$-dimensional parameter space. Instead, the algorithm requires only:

\begin{itemize}
  \item generating samples in the current state $ |\psi_t\rangle $ to estimate $ A_t $;  
  \item applying $\textcolor{black}{\hat U_d(\beta_{t+1}\Delta t)\hat U_p(\Delta t)}$, with the rotation angle calibrated directly by the measurement.
\end{itemize}

Although FALQON requires, at each iteration, only the estimation of the expectation value of a single operator, this operator corresponds to the commutator ($i[\hat H_d,\hat H_p]$), which is generally non-diagonal. As a result, its estimation requires measurements in multiple bases, in contrast to the cost energy associated with $\hat H_p$, which can be obtained directly in the computational basis. This introduces an additional experimental overhead that is absent in a simple cost function evaluation.

Furthermore, FALQON follows an iterative structure in which a new circuit layer is appended at each step of the algorithm, leading to a monotonic increase in circuit depth with the number of iterations. Numerical studies have shown that, for several problem instances, a large number of layers may be required to achieve convergence, resulting in circuits that are significantly deeper than those typically considered in QAOA with a fixed number of layers \cite{Larsen2024,Pexe2024}. This growing circuit depth represents a practical limitation in noisy intermediate scale quantum devices and is one of the main factors motivating the development of modified and extended variants of the original FALQON protocol.

\subsection{Warm-starting and warm-start-driven mixer}\label{subsec:warm_start_ws}

To accelerate convergence, we prepare a coherent product state biased towards a good classical approximation.  
Following \cite{Egger2021}, we first solve a continuous relaxation of the discrete portfolio problem, yielding a vector $\boldsymbol{c}\in[0,1]^n$ of fractional inclusion levels that satisfies the budget constraint $\sum_{i=1}^n c_i = K$. Interpreting $c_i$ as an inclusion probability, we clip the entries to avoid extremal angles,
\[
p_i=\min\{\max\{c_i,\epsilon\},1-\epsilon\}, \qquad \epsilon\in[0,1/2),
\]
and define
\[
\theta_i=\arcsin\sqrt{p_i}.
\]
Starting from $\ket{0}^{\otimes n}$, we apply a single-qubit rotation $R_Y(2\theta_i)$ to qubit $i$, obtaining
\[
\begin{aligned}
    \ket{\phi_0}
& =\bigotimes_{i=1}^{n}\Bigl(\cos\theta_i\,\ket{0}
                     +\sin\theta_i\,\ket{1}\Bigr) \\
& = \bigotimes_{i=1}^{n}\Bigl(\sqrt{1-p_i}\,\ket{0}
                     +\sqrt{p_i}\,\ket{1}\Bigr).
\end{aligned}
\]
This construction encodes the relaxed solution in the measurement statistics: for each qubit,
$\Pr(1)=|\sin\theta_i|^2=\sin^2\theta_i=p_i$, so that samples are biased towards configurations consistent with the relaxation. The clipping parameter $\epsilon$ prevents $p_i\in\{0,1\}$, which would freeze qubits into deterministic values and can reduce the ability of the subsequent dynamics to explore nearby configurations. Angle damping $\theta_i\mapsto \lambda\theta_i$ with $0<\lambda<1$ is a possible refinement, but it is not used in the experiments reported here.

Besides using $\ket{\phi_0}$ as an initial state, we also employ a single-qubit driver Hamiltonian whose local ground state matches the warm-start amplitudes. Using the same angles $\theta_i=\arcsin\sqrt{p_i}$ defined above, we set

\begingroup
\color{black}
\begin{equation}
\hat H_{ws}
= -\sum_{i=1}^{n}\bigl(\sin\theta_i\,X_i + \cos\theta_i\,Z_i\bigr).
\end{equation}
\endgroup

This choice aligns each qubit's effective field with the Bloch vector of the warm-start state and biases the evolution towards the relaxed solution. Unlike the XY mixer, $\hat H_{ws}$ does not conserve Hamming weight; therefore, in this work it is treated as a distinct mixer (separate from the XY setting), consistent with the three-mixer comparison reported in Sec.~\ref{sec:results}.

\subsection{Transverse \texorpdfstring{$X$}{X} mixer}\label{subsec:mixerX}

When using the transverse mixer, we adopt the uniform superposition $\ket{+}^{\otimes n}$, which is an eigenstate of $\hat H_x$.

\begingroup
\color{black}
\begin{equation}
\hat H_x = \sum_{i=1}^{n} X_i.
\end{equation}
\endgroup

\textcolor{black}{
The transverse mixer generates a quantum walk on the $n$-dimensional hypercube, whose $2^n$ vertices correspond to the computational basis states. Each Pauli-$X$ operator flips a single qubit, so every basis state is connected only to states that differ by one bit, i.e., at Hamming distance one \cite{Nzongani2025}. Consequently, the evolution can explore the entire configuration space through successive single-bit flips. This mixer is appropriate when the feasible set imposes no cardinality constraints. However, because the hypercube contains both feasible and infeasible configurations, the dynamics also transfers amplitude to inadmissible states whenever constraints are present \cite{Magann2022, Hadfield2017_QAOAplus, Fuchs2022_ConstrainedMixers}.
}


\subsection{XY mixer with cardinality conservation}\label{subsec:mixerXY}

For the XY setting, we initialize the dynamics in the symmetric Dicke state with $K$ excitations, i.e., the uniform superposition over all bitstrings of Hamming weight $K$:
\[
\ket{D^{(n)}_{K}}
=
\frac{1}{\sqrt{\binom{n}{K}}}
\sum_{\substack{z\in\{0,1\}^n\\ |z|=K}}
\ket{z},
\]

where $|z|=\sum_{i=1}^n z_i$. This choice ensures that the subsequent XY evolution remains within the feasible $K$-excitation subspace.

\begingroup
\color{black}
\begin{equation}
\hat{H}_{xy}
=
\frac{1}{2}
\sum_{i<j}
\left(X_iX_j+Y_iY_j\right).
\end{equation}
\endgroup

This operator preserves the total number of excitations, remaining within a fixed-Hamming-weight subspace.  
Such a feature makes it compatible with problems requiring $\sum_i x_i = K$. The XY mixer has also been shown to induce optimization landscapes more favorable than the conventional $X$ mixer in constrained settings \cite{Wang2020XY, Hadfield2017_QAOAplus, Fuchs2022_ConstrainedMixers}. In our simulations we employ an all-to-all coupling (complete graph), i.e., the sum runs over all pairs $i<j$.

\section{Variants}\label{sec:trso}

This section summarizes the two FALQON variants used in this work, namely TR-FALQON and SO-FALQON.
In both cases we preserve the same layer structure described in Sec.~\ref{sec:falqon}; the only change
is the feedback rule used to compute the control parameter $\beta_k$.

\subsection{TR-FALQON: time-rescaled feedback optimization}\label{subsec:trfalqon}

{\color{black}
Time-Rescaled FALQON is a variant of FALQON that incorporates the time-rescaling method into the reference dynamics of the standard FALQON. Before presenting the algorithm itself, it is convenient to discuss this method.

The time-rescaling method is a strategy introduced in Ref. \cite{Bernardo2020}. Consider a reference dynamics governed by a time-dependent Hamiltonian $\hat{H}(t)$, which drives the system from an initial state $\ket{\psi_i}$ to a final state $\ket{\psi_f}$ over a time interval ranging from $t_i=0$ to $t_f$. This dynamics is described by the time-evolution operator 

{\color{blue}
\begin{equation}
\hat{U}(t_f) = \hat{\mathcal{T}} \exp\left\{-i\int_0^{t_f} \hat{H}\left(t^\prime\right) d t^\prime\right\},
\end{equation}}
where $\hat{\mathcal{T}}$ denotes the time-ordering operator. If we perform the time transformation $t=f(\tau)$, the time-evolution operator can be rewritten as
{\color{blue}
\begin{equation}
\hat{U}\left(f^{-1}(t_f)\right) = \hat{\mathcal{T}} \exp\left\{-i\int_{f^{-1}(0)}^{f^{-1}(t_f)} \hat{H}\left(f(\tau)\right)\dot{f}(\tau), d\tau\right\}.
\end{equation}}

The function $f(\tau)$ is referred to as the time-rescaling function, while $f^{-1}$ and $\dot{f}$ denote its inverse and time derivative, respectively. Defining

\begin{equation}
\hat{\mathcal{H}}(\tau)=\hat{H}\left(f(\tau)\right)\dot{f}(\tau),
\end{equation}
and choosing $f$ such that $f^{-1}(0)=0$, it follows that the dynamics governed by $\hat{\mathcal{H}}$, hereafter referred to as the time-rescaled dynamics, drives the system from $\ket{\psi_i}$ to $\ket{\psi_f}$ in a time interval $\Delta_t=f^{-1}(t_f)$. Therefore, whenever $f^{-1}(t_f)<t_f$, the same process is completed in a shorter time than that required by the reference dynamics.

When the reference dynamics describes an adiabatic process, the time-rescaled dynamics constitutes a shortcut to adiabaticity, provided that the function $f$ satisfies the following conditions

\begin{enumerate}
\item $f^{-1}(0)=0$,
\item $f^{-1}(t_f)<t_f$,
\item $\dot{f}(0)=\dot{f}(f^{-1}(t_f))=1$.
\end{enumerate}

The first two conditions ensure that the evolution starts at the same initial time as the reference dynamics and finishes in a shorter time interval. The third condition guarantees that the Hamiltonians $\hat{H}$ and $\hat{\mathcal{H}}$ coincide at the beginning and at the end of the evolution.

A function satisfying these conditions was proposed in Ref. \cite{Bernardo2020},

\begin{equation}
f(t)=a t-\frac{(a-1)}{2 \pi a} t_f \sin \left(\frac{2 \pi a}{t_f} t\right),
\label{eq:reescaled_function}
\end{equation}
and is commonly adopted as the standard choice in works employing the time-rescaling method. It was used in Ref. \cite{Ferreira2026}, which showed that the shortcut to adiabaticity obtained through time rescaling follows the same sequence of states as the reference adiabatic dynamics. The same function was also adopted in the work that introduced TR-FALQON \cite{Rattighieri2025}. For this choice of $f$, it follows that $f^{-1}(t_f)=t_f/a$. For this reason, the parameter $a$ is referred to as the time-contraction parameter.

Thus, TR-FALQON rescales the dynamics generated by the standard FALQON in order to reduce the convergence time. By employing the rescaled Hamiltonian $\hat{\mathcal{H}}(\tau)$, the evolution operators of each layer become the standard FALQON evolution operators parametrized by the derivative of the time-rescaling function. As a result, the effective time step of the $k$th layer becomes $\dot{f}(\tau_k)\Delta t$. In our implementation, $\dot{f}(\tau)$ is the derivative of the function defined in Eq. \eqref{eq:reescaled_function}, given by

\begin{equation}\label{eq:tr_fprime}
\dot{f}(\tau)=a-(a-1)\cos\left(\frac{2\pi a}{t_f}\tau\right),
\qquad
a\ge1, \qquad
t_f>0, \quad
\tau\in[0,t_f].
\end{equation}

Denoting $\tau_k=k\Delta t$, one TR-FALQON layer is implemented as

\begin{equation}\label{eq:tr_layer}
\ket{\psi_{k+1}}= \hat U_d\left(\dot{f}(\tau_k)\beta_k\Delta t\right)
\hat U_p\left(\dot{f}(\tau_k)\Delta t\right)
\ket{\psi_k}.
\end{equation}

The feedback parameter is updated according to the same commutator-based rule used in FALQON. The only difference is that it is rescaled by the derivative of the time-rescaling function evaluated at the next time step, yielding

\begin{equation}\label{eq:tr_beta}
\beta_{k+1}= -\frac{A_k}{\dot{f}(\tau_{k+1})}
-\frac{\bigl\langle \psi_{k}\bigm|, i[\hat H_d,\hat H_p] ,\bigm|\psi_{k}\bigr\rangle}{\dot{f}(\tau_{k+1})}.
\end{equation}
}

\subsection{SO-FALQON: second-order feedback optimization}
\label{subsec:sofalqon}

\textcolor{black}{Second-order FALQON refines the feedback update by incorporating second-order information via a quadratic approximation \cite{Arai2025}. The first-order candidate $\beta^{(1)}$ follows directly from the standard FALQON feedback rule defined in Eq.~\eqref{eq:lei de feedback1}. In SO-FALQON, two additional scalar quantities are introduced,}

\begingroup
\color{black}
\begin{equation}
B =
\bigl\langle \psi \bigm|\,
\frac{1}{2}
[[\hat H_d,\hat H_p],\hat H_d]
\,\bigm|\psi\bigr\rangle,
\label{eq:so_B_definition}
\end{equation}

\begin{equation}
C =
\bigl\langle \psi \bigm|\,
[[\hat H_d,\hat H_p],\hat H_p]
\,\bigm|\psi\bigr\rangle.
\label{eq:so_C_definition}
\end{equation}
\endgroup

\textcolor{black}{These quantities encode the second-order corrections. To derive the second-order candidate, we consider the change in the expected cost energy produced by one layer,}

\begingroup
\color{black}
\begin{equation}
\begin{split}
\Delta E(\beta)
&= 
\bigl\langle\psi\bigm|
\hat U_p^{\dagger}(\Delta t)
\hat U_d^{\dagger}(\beta\Delta t)
\hat H_p
\hat U_d(\beta\Delta t)
\hat U_p (\Delta t)
\bigm|\psi\bigr\rangle \\
&\quad-
\bigl\langle\psi\bigm|
\hat H_p
\bigm|\psi\bigr\rangle.
\end{split}
\label{eq:so_energy_difference}
\end{equation}
\endgroup

\textcolor{black}{Expanding the evolution operators up to second order in $\Delta t$,}

\begingroup
\color{black}
\begin{align}
\hat U_d(\beta \Delta t)
&\simeq
\mathbb{I}
-i\beta\hat H_d\Delta t
-\frac{1}{2}
\left(\beta\hat H_d\Delta t\right)^2,
\\
\hat U_p(\Delta t)
&\simeq
\mathbb{I}
-i\hat H_p\Delta t
-\frac{1}{2}
\left(\hat H_p\Delta t\right)^2,
\end{align}
\endgroup

\textcolor{black}{the energy variation becomes}

\begingroup
\color{black}
\begin{equation}
\Delta E(\beta)
\simeq
\Delta t\,\beta A
+
(\Delta t)^2\beta^2 B
+
(\Delta t)^2\beta C
+
\mathcal{O}\!\left((\Delta t)^3\right).
\label{eq:so_energy_expansion}
\end{equation}
\endgroup

\textcolor{black}{This expression can be written as a quadratic function of $\beta$,}

\begingroup
\color{black}
\begin{equation}
\begin{split}
\Delta E(\beta)
&\simeq
(\Delta t)^2 B
\left(
\beta+
\frac{A+\Delta t\,C}
     {2\Delta t\,B}
\right)^2 \\
&\quad-
\frac{\left(A+\Delta t\,C\right)^2}{4B}
+
\mathcal{O}\!\left((\Delta t)^3\right).
\end{split}
\label{eq:so_completed_square}
\end{equation}
\endgroup

\textcolor{black}{For $B>0$, minimizing this quadratic approximation gives the second-order candidate. For $B<0$, the sign is reversed to preserve a negative energy variation. Both cases can be expressed compactly using $\lvert B\rvert$:}

\begin{equation}\label{eq:so_beta2}
\beta^{(2)}
=
-\frac{A + \Delta t\, C}{2\,\Delta t\, |B|}.
\end{equation}

In our implementation, the second-order scalars are computed as

\begin{equation}\label{eq:so_B}
\begin{split}
B
&=
\bigl\langle \psi \bigm|
\hat H_d \hat H_p \hat H_d
\bigm|\psi \bigr\rangle \\
&\quad-\frac{1}{2}\Bigl(
\bigl\langle \psi \bigm|
\hat H_d \hat H_d \hat H_p
\bigm|\psi \bigr\rangle
+
\bigl\langle \psi \bigm|
\hat H_p \hat H_d \hat H_d
\bigm|\psi \bigr\rangle
\Bigr),
\end{split}
\end{equation}

\begin{equation}\label{eq:so_C}
\begin{split}
C
&=
-2\bigl\langle \psi \bigm|
\hat H_p \hat H_d \hat H_p
\bigm|\psi \bigr\rangle \\
&\quad+
\bigl\langle \psi \bigm|
\hat H_p \hat H_p \hat H_d
\bigm|\psi \bigr\rangle
+
\bigl\langle \psi \bigm|
\hat H_d \hat H_p \hat H_p
\bigm|\psi \bigr\rangle.
\end{split}
\end{equation}

Finally, the hybrid strategy used here selects between the first- and second-order candidates as

\begin{equation}\label{eq:so_hybrid}
\beta =
\begin{cases}
\beta^{(2)}, & |\beta^{(1)}| > |\beta^{(2)}|,\\
\beta^{(1)}, & \text{otherwise}.
\end{cases}
\end{equation}

This choice is made to prevent $\beta^{(2)}$ from breaking convergence due to an excessively large update. With this modified feedback rule, larger time steps can be employed without compromising convergence.

\section{Implementation}\label{sec:implementation}

All experiments were implemented in Python using statevector simulation with complex amplitudes in double precision. The portfolio instances and their financial statistics were generated with the Qiskit Finance data provider \cite{JavadiAbhari2024_Qiskit, QiskitFinance_Tutorial_PortfolioOptimization}, while the quantum time evolutions, feedback rules, and state-preparation utilities were executed with the open-source \texttt{qtsuru} framework; in particular, the state constructors used throughout this work follow the implementations provided in \cite{qsim_states}.

For the main study we consider $n=6$ assets and a fixed budget $K=B=3$. \textcolor{black}{To instantiate this setting, we generated a synthetic portfolio instance using \texttt{RandomDataProvider} from \texttt{qiskit\_finance.data\_providers} \cite{QiskitFinance_Tutorial_PortfolioOptimization} over the fixed time window from January 1, 2016, to January 30, 2016, with $25$ random seeds (all the details are available in the \cite{FALQONPortfolio}). This provider generates reproducible mock stock-price trajectories by applying pseudo-random Gaussian increments to each asset path. It was chosen because it yields benchmark instances that are independent of external market-data APIs and fully reproducible from the reported seed and time window \cite{he2023alignment, sureshbabu2024parameter}. From the resulting price trajectories, we computed the period returns and estimated the mean return vector $\boldsymbol{\mu}$ and covariance matrix $\boldsymbol{\Sigma}$.} Given $(\boldsymbol{\mu},\boldsymbol{\Sigma})$, we built the problem Hamiltonian $\hat H_p$ as a diagonal operator whose computational-basis eigenvalues match the QUBO objective, including the quadratic cardinality penalty. In our implementation, the diagonal energies are computed explicitly for all $2^n$ basis states, allowing us to identify the exact ground-state bitstring(s) by exhaustive search and to track their total probability mass during the quantum evolution.

We compared three driver constructions (mixers), each paired with the feedback-based optimizers considered in this work. For the transverse mixer we use $\hat H_x$ and initialize the dynamics from the uniform superposition. For the cardinality-preserving XY mixer we use $\hat H_{xy}$ and initialize from a fixed-Hamming-weight state of weight $B$, ensuring that the evolution remains in the feasible subspace. For warm-starting we first solve the continuous relaxation of the portfolio problem under the same budget constraint, obtaining a vector $\boldsymbol{c}^\star\in[0,1]^n$. We then prepare a product state by applying single-qubit $R_y$ rotations whose angles are determined by $\boldsymbol{c}^\star$ with $\epsilon$-clipping, producing an initial state biased toward the relaxed solution. When using the warm-start driver, the mixer itself is constructed from these angles and applied via a sequence of single-qubit rotations consistent with the \texttt{qsim} implementation.

Each run \textcolor{black}{records the layer-by-layer values of} the expected cost energy $\langle \hat H_p\rangle_k$, the feedback control values $\beta_k$, and a success metric $P_k$, defined as the total probability on the exact ground-state configuration(s) of $\hat H_p$. \textcolor{black}{To assess robustness in the $n=6$ benchmark, each configuration was executed using 25 random seeds. The reported curves correspond to the per-layer averages of the energy and success probability over these 25 runs.} The mean energy and mean success-probability \textcolor{black}{curves} were computed by averaging the corresponding per-layer values across seeds, yielding the final figures used for comparison among mixers and algorithmic variants.

In addition to the main $n=6$ study, we carried out a secondary experiment with $n=10$ assets, \textcolor{black}{using the same fixed time window as before}, to probe how the different mixers and feedback variants concentrate probability mass onto the optimal configuration. The goal of this larger instance is not statistical benchmarking, but rather a qualitative assessment of convergence behavior toward a regime where the optimal bitstring becomes unambiguously dominant; accordingly, we track whether the success probability exceeds a $50\%$ threshold, at which point the resulting state is clearly separated from the remaining configurations.

\section{Results}\label{sec:results}


For completeness, the success probability used in the analysis is written explicitly as

\begingroup
\color{black}
\begin{equation}
P_k=\sum_{z\in\mathcal{G}}|\langle z|\psi_k\rangle|^2,
\end{equation}
\endgroup
where $\mathcal{G}$ is the set of exact ground-state bitstrings of $\hat H_p$ (degeneracies, when present, are included in the sum).

\begin{figure*}[t]
\centering
\includegraphics[width=\textwidth]{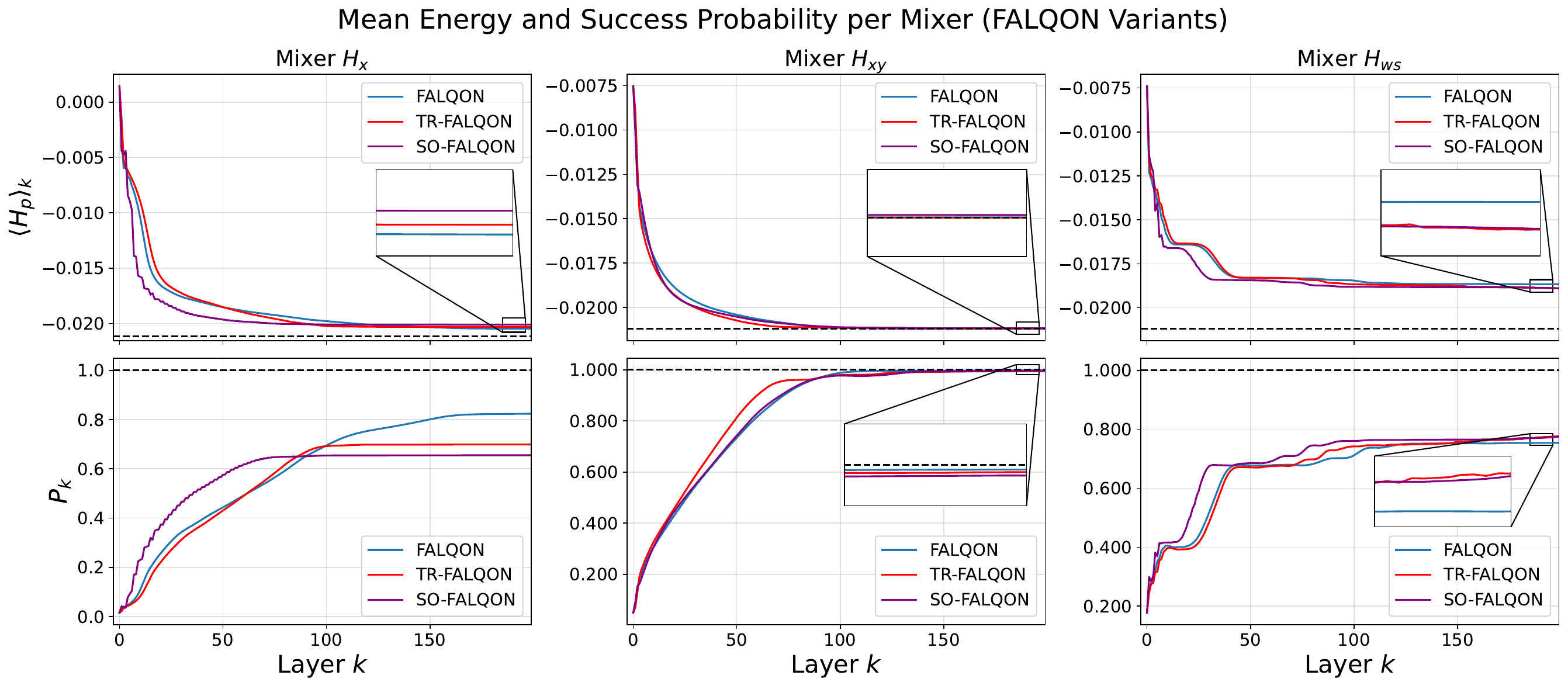}
\caption{\textcolor{black}{Mean energy (top row) and success probability (bottom row) versus layer $k$, averaged over 25 seeds for $n=6$. Columns correspond to the three mixers $\hat H_x$, $\hat H_{xy}$, and $\hat H_{ws}$. The dashed line in the energy panels marks the exact ground-state energy of $\hat H_p$. Zoomed insets highlight regions where the curves nearly overlap, allowing a more detailed comparison between the algorithmic variants.}}
\label{fig:mean_mixers}
\end{figure*}

Figure~\ref{fig:mean_mixers} summarizes the mean dynamics for all mixers and algorithmic variants, with the dashed line indicating the exact ground-state energy. To facilitate a direct comparison at fixed depths, Table~\ref{tab:mixers_checkpoints} reports $P_k$ at $k\in\{25,50,75,100\}$.

\begin{table}[h]
\centering
\footnotesize
\setlength{\tabcolsep}{4pt}
\caption{Success probability $P_k$ (mean over 25 seeds) at selected depths for each optimizer--mixer configuration ($n=6$). The best value in each column is highlighted.}
\label{tab:mixers_checkpoints}
\begin{tabular}{llcccc}
\hline
Mixer & Variant & $k=25$ & $k=50$ & $k=75$ & $k=100$ \\
\hline
$\hat H_x$ & FALQON & 30.56\% & 44.32\% & 56.52\% & 69.49\% \\
$\hat H_x$ & TR-FALQON & 27.03\% & 43.02\% & 58.31\% & 69.21\% \\
$\hat H_x$ & SO-FALQON & 41.39\% & 57.60\% & 64.69\% & 65.42\% \\
\hline
$\hat H_{xy}$ & FALQON & 49.03\% & 73.37\% & 91.24\% & \textbf{98.80\%} \\
$\hat H_{xy}$ & TR-FALQON & 52.28\% & \textbf{81.31\%} & \textbf{95.93\%} & 97.97\% \\
$\hat H_{xy}$ & SO-FALQON & 50.08\% & 74.23\% & 91.87\% & 97.71\% \\
\hline
$\hat H_{ws}$ & FALQON & 43.88\% & 67.75\% & 67.75\% & 71.15\% \\
$\hat H_{ws}$ & TR-FALQON & 41.60\% & 67.08\% & 69.21\% & 74.20\% \\
$\hat H_{ws}$ & SO-FALQON & \textbf{59.10\%} & 68.55\% & 72.24\% & 76.05\% \\
\hline
\end{tabular}
\end{table}

\noindent
\textbf{Transverse mixer $\boldsymbol{\hat H_x}$:}
All three feedback rules drive the mean energy close to the optimal value, but their probability concentration differs. As shown in Fig.~\ref{fig:mean_mixers} and Table~\ref{tab:mixers_checkpoints}, SO-FALQON yields the fastest initial decrease of $\langle \hat H_p\rangle_k$ and also attains the highest $P_k$ at early depths ($k=25,50,75$). However, its probability growth saturates earlier, while standard FALQON continues to concentrate probability mass and becomes the best-performing variant by $k=100$, with TR-FALQON exhibiting intermediate performance. Although the energy curves nearly overlap at larger depths, the corresponding success probabilities remain distinct. This behavior suggests that the first-order Trotter approximation has a stronger impact on probability concentration than on the average energy. Since standard FALQON evolves with the nominal time step $\Delta t$, it is expected to accumulate smaller discretization errors than TR-FALQON, whose effective time step is $\dot{f}(\tau_k)\Delta t$, and SO-FALQON, whose feedback law is not derived directly from the Lyapunov control formulation. Consequently, FALQON continues to increase the probability of the optimal state over a larger number of layers, whereas TR-FALQON and SO-FALQON approach a plateau earlier. This behavior is also consistent with the fact that the transverse mixer explores the full hypercube and does not preserve the budget subspace. As a result, even when the mean energy is close to the optimum, amplitude can remain distributed among near-optimal or constraint-violating configurations, limiting the probability accumulated on the exact ground state.

\noindent
\textcolor{black}{\textbf{XY mixer $\boldsymbol{\hat H_{xy}}$:}}
The XY mixer yields the most pronounced probability concentration. All three variants rapidly increase $P_k$ and converge essentially to the ground-state energy, with $P_k$ already above $98\%$ by $k=75$ for all feedback rules (Table~\ref{tab:mixers_checkpoints}) and approaching unity thereafter (Fig.~\ref{fig:mean_mixers}). In this regime, TR-FALQON tends to reach the high-probability plateau slightly earlier at intermediate depths (e.g., $k=50$), while the overall behavior remains similar across variants once the dynamics enters the feasible Hamming-weight sector. This result highlights the practical advantage of enforcing the cardinality constraint at the driver level: by restricting evolution to the fixed-excitation subspace, the algorithm avoids wasting amplitude on infeasible strings and makes the optimal bitstring dominant within comparatively few layers.

\noindent
\textcolor{black}{\textbf{Warm-start-driven mixer $\boldsymbol{\hat H_{ws}}$:}}
The warm-start-driven configuration displays a characteristic step-like growth in success probability, with a rapid initial rise followed by slower improvements. This behavior is expected when the initialization and the driver embed prior information from the continuous relaxation: the evolution begins closer to low-energy regions, yielding an early gain in $P_k$. However, similarly to the transverse mixer $\hat H_x$, the warm-start driver $\hat H_{ws}$ does not preserve the subspace that satisfies the budget constraint. Consequently, the evolution can also populate infeasible bitstrings that violate the cardinality constraint, reducing the probability concentration on the exact optimal solution. As a result, the final success probabilities remain below the near-deterministic behavior observed with the XY mixer, while the energy approaches the optimum more gradually. At the checkpoint depths reported in Table~\ref{tab:mixers_checkpoints}, SO-FALQON yields the largest $P_k$ across all four depths; at larger depths the three curves become closer, indicating that \textcolor{black}{$\hat H_{ws}$} provides robust early-stage acceleration but can stabilize around a broader neighborhood of competitive configurations rather than concentrating all amplitude onto the exact optimum as strongly as \textcolor{black}{$\hat H_{xy}$}.

\noindent
Overall, Fig.~\ref{fig:mean_mixers} indicates that mixer structure is a dominant factor for solution visibility: while all variants can approach the optimal energy, achieving a clearly dominant optimal bitstring (high $P_k$) is markedly easier when the driver respects the problem constraint (XY) or embeds a strong prior (warm-start), with the XY mixer producing the most decisive concentration in the $n=6$ benchmark.

\noindent
\textbf{Secondary experiment ($n=10$) and solution visibility.}


\begin{figure*}[!t]
\centering
\includegraphics[width=\textwidth]{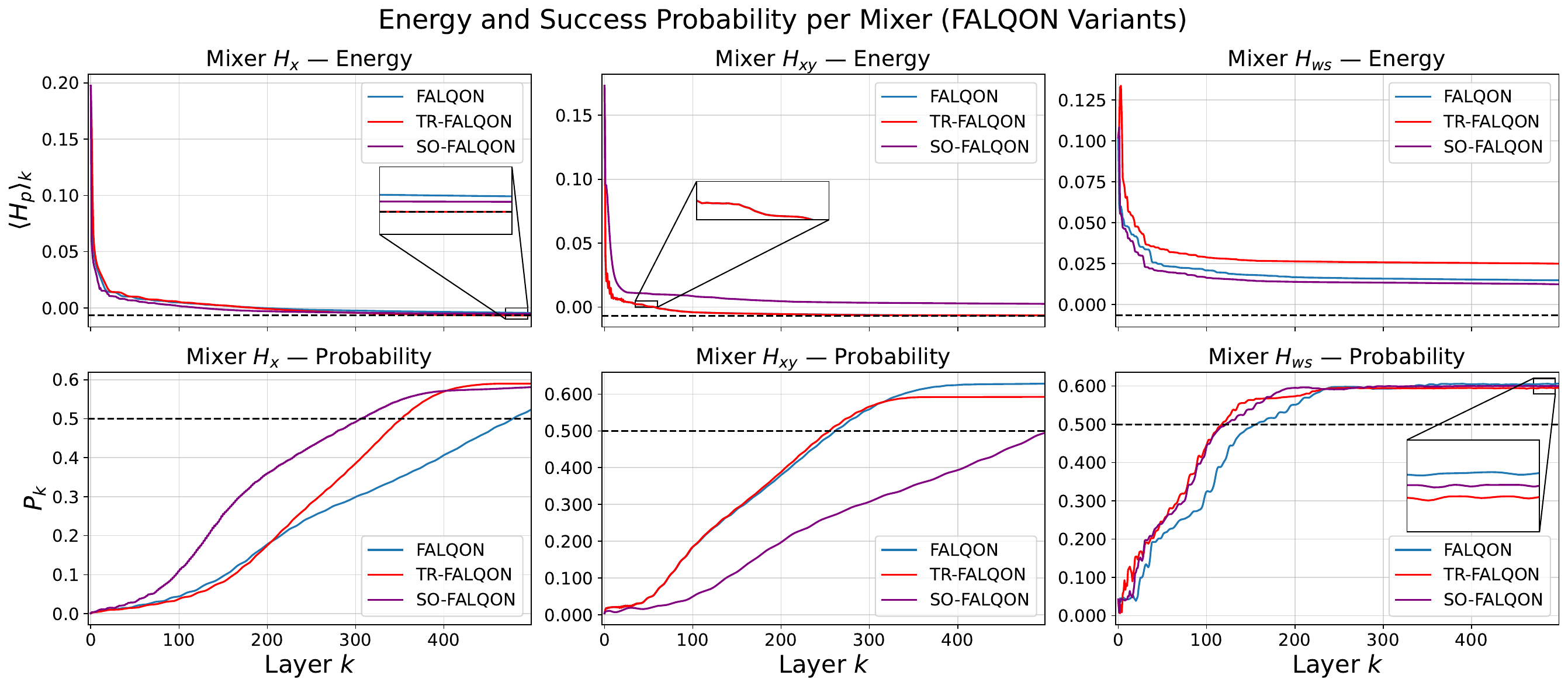}
\caption{Energy (top row) and success probability (bottom row) versus layer $k$ for $n=10$ (single run). Columns correspond to the mixers \textcolor{black}{$\hat H_x$}, \textcolor{black}{$\hat H_{xy}$}, and \textcolor{black}{$\hat H_{ws}$}. Dashed lines indicate the exact ground-state energy (top) and \textcolor{black}{$P_k=0.5$} (bottom). The insets focus on regions where the curves are nearly indistinguishable, making the differences between the algorithmic variants easier to observe.}
\label{fig:mixers_10assets}
\end{figure*}

Figure~\ref{fig:mixers_10assets} reports the layer-by-layer evolution of the expected energy and success probability under the same three mixers.


\textcolor{black}{As defined in Sec.~\ref{sec:implementation}, we use $P_k>0.5$ as the solution-visibility threshold.}

\begin{figure*}[t]
\centering
\includegraphics[width=\textwidth]{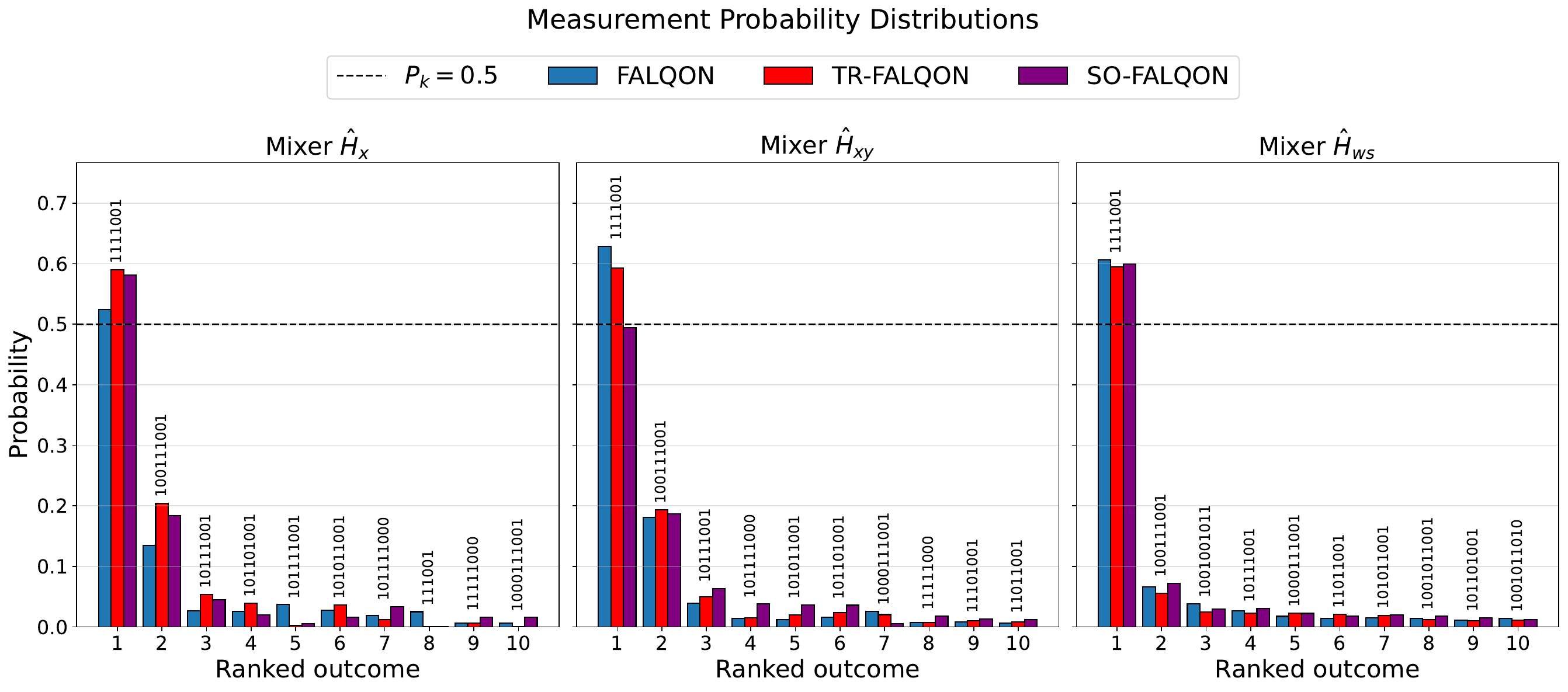}
\caption{\textcolor{black}{Final measurement-probability distributions for the $n=10$ instance after 500 layers. Columns correspond to the transverse $\hat H_x$, XY $\hat H_{xy}$, and warm-start $\hat H_{ws}$ mixers. For each ranked outcome, the black, red, and purple bars represent FALQON, TR-FALQON, and SO-FALQON, respectively. Rank 1 denotes the total ground-state probability $P_k$, whereas ranks 2--10 correspond to the nine non-ground-state bitstrings with the highest probabilities across the three variants for each mixer. The dashed line indicates the solution-visibility threshold $P_k=0.5$.}}
\label{fig:measurement_distribution}
\end{figure*}

\textcolor{black}{To illustrate this criterion, Fig.~\ref{fig:measurement_distribution} presents the final measurement-probability distributions after 500 layers. In each panel, rank 1 represents the total ground-state probability $P_k$, whereas ranks 2--10 represent the nine most relevant non-ground-state bitstrings across the three feedback variants. When $P_k>0.5$, the ground-state configuration accumulates more probability than all remaining outcomes combined, making the optimal portfolio clearly identifiable through measurement.}

\noindent
\textcolor{black}{\textbf{Transverse mixer $\boldsymbol{\hat H_x}$:}}
All variants decrease the energy steadily, but probability concentration is slow: within the first 200 layers, $P_k$ remains well below the $50\%$ visibility threshold (Table~\ref{tab:mixers10_checkpoints}). This is consistent with unconstrained exploration of the full hypercube, where amplitude can be spread across many near-optimal configurations. In this regime, lowering $\textcolor{black}{\langle \hat H_p\rangle_k}$ does not immediately translate into a clearly dominant optimal bitstring, and the depth required to surpass $P_k>0.5$ can become large.

\begin{table}[t]
\centering
\footnotesize
\setlength{\tabcolsep}{4pt}
\caption{Checkpoint success probability $P_k$ for $n=10$ (single run) at selected depths for each optimizer--mixer configuration. For each depth, the best value is highlighted.}
\label{tab:mixers10_checkpoints}
\begin{tabular}{llcccc}
\hline
Mixer & Variant & $k=50$ & $k=100$ & $k=150$ & $k=200$ \\
\hline
$\hat H_x$ & FALQON     & 4.11\%  & 6.80\%  & 10.63\% & 14.41\% \\
$\hat H_x$ & TR-FALQON  & 2.51\%  & 4.87\%  & 9.19\%  & 17.16\% \\
$\hat H_x$ & SO-FALQON  & 6.16\%  & 9.94\%  & 13.98\% & 17.28\% \\
\hline
$\hat H_{xy}$ & FALQON     & 10.67\% & 18.68\% & 30.02\% & 38.52\% \\
$\hat H_{xy}$ & TR-FALQON  & 10.80\% & 18.80\% & 30.19\% & 39.11\% \\
$\hat H_{xy}$ & SO-FALQON  & 2.03\%  & 5.34\%  & 14.13\% & 23.42\% \\
\hline
$\hat H_{ws}$ & FALQON     & 13.44\% & 38.57\% & 56.19\% & 59.45\% \\
$\hat H_{ws}$ & TR-FALQON  & \textbf{14.07\%} & \textbf{45.32\%} & \textbf{58.96\%} & 59.47\% \\
$\hat H_{ws}$ & SO-FALQON  & 10.47\% & 35.49\% & 53.96\% & \textbf{59.76\%} \\
\hline
\end{tabular}
\end{table}

\noindent
\textcolor{black}{\textbf{XY mixer $\boldsymbol{\hat H_{xy}}$:}}
The excitation-preserving mixer improves energy convergence (notably for FALQON and TR-FALQON), yet probability growth remains gradual in this $n=10$ instance. While \textcolor{black}{$\hat H_{xy}$} restricts the dynamics to a fixed-Hamming-weight sector, the visibility threshold may still require more depth than in the $n=6$ case, indicating that constraint-preserving dynamics alone is not sufficient to guarantee early dominance of the exact optimum as the instance size increases.

\noindent
\textcolor{black}{\textbf{Warm-start-driven mixer $\boldsymbol{\hat H_{ws}}$:}}
The warm-start-driven mixer provides the earliest route to solution visibility: $P_k$ rises quickly and crosses the $50\%$ threshold substantially earlier than with \textcolor{black}{$\hat H_x$} and \textcolor{black}{$\hat H_{xy}$}, reaching $\approx 0.59\!-\!0.60$ by 150--200 layers (Table~\ref{tab:mixers10_checkpoints}). This behavior is aligned with the role of warm-start information as a structured prior: the dynamics begins in a region already biased toward competitive configurations and can accumulate probability mass on the optimum faster, even when the energy convergence is not yet fully saturated.

In this work, we deliberately employ a relatively large time step $\Delta t$ to accelerate convergence in terms of circuit depth. This choice was motivated by our objective of emphasizing the differences between the proposed feedback strategies and mixer constructions within a practical number of layers. Using a smaller $\Delta t$ would require substantially deeper circuits before noticeable differences in convergence behavior become apparent, making the comparison less informative for near-term quantum devices.

The use of a larger time step, however, increases the error associated with the first-order Trotter approximation employed by FALQON, TR-FALQON, and SO-FALQON. As $\Delta t$ increases, the implemented discrete evolution progressively departs from the continuous-time dynamics assumed in the Lyapunov-based derivation of the feedback laws. As a result, the saturation observed in the energy and success-probability curves can be partially attributed to this deviation.

The impact of this approximation is not the same for the three algorithms. Standard FALQON is less affected, since its evolution is governed directly by the nominal time step $\Delta t$, maintaining a closer correspondence with the original Lyapunov dynamics. In contrast, TR-FALQON evolves with the effective time step $\dot{f}(\tau_k)\Delta t$, which can become larger than the nominal value during the evolution, increasing discretization errors. In SO-FALQON, the feedback parameter is obtained from a second-order approximation of the energy variation rather than from the original Lyapunov feedback law. This strategy enables a faster initial decrease in the objective energy, but the implemented dynamics tends to depart from the continuous-time evolution as higher-order terms become more relevant.

The comparison between mixers indicates, however, that Trotterization is not the only mechanism responsible for the observed behavior. The $n=6$ benchmark shows that the dominant effect is the choice of mixer. Unlike $\hat H_{xy}$, both $\hat H_x$ and $\hat H_{ws}$ do not preserve the fixed-cardinality subspace, allowing population transfer to infeasible bitstrings throughout the evolution. As a result, probability remains distributed among feasible and infeasible configurations, limiting concentration on the exact optimum even when the average energy is already close to its minimum. For the larger $n=10$ instance, the effects associated with the discrete implementation become more evident, suggesting that Trotterization errors become progressively more relevant as the problem size increases.

From an applications perspective, the $P_k>0.5$ criterion is directly related to the practical use of quantum portfolio optimization. Portfolio optimization is typically repeated under changing market conditions, such as portfolio rebalancing, stress testing, and scenario analysis. A method that makes the optimal portfolio \emph{visible} with fewer layers can therefore reduce both circuit depth and the number of measurement shots required to identify the correct allocation \cite{Sakuler2023_RealWorldQA_Portfolio}. This consideration is particularly relevant in the NISQ regime, where coherence times and two-qubit gate noise impose strict depth limitations. The present results suggest that the choice of hyperparameters, including $\Delta t$, the time-rescaling function in TR-FALQON, and the warm-start parameters of $\hat H_{ws}$, should balance two competing objectives: accelerating the initial energy reduction while delaying the saturation of the success probability.

\section{Conclusion}\label{sec:conclusion}

We benchmarked nine configurations obtained by combining three feedback-based optimizers with three driver Hamiltonians for discrete, cardinality-constrained portfolio optimization. The results show that, while all variants can reduce the expected energy toward the ground-state value, the mixer structure is the dominant factor for making the optimal portfolio \emph{visible} in measurements: the excitation-preserving \textcolor{black}{$\hat H_{xy}$} consistently concentrates probability on the ground-state manifold more decisively than the transverse \textcolor{black}{$\hat H_x$}, whereas the warm-start-driven \textcolor{black}{$\hat H_{ws}$} accelerates early progress by biasing the evolution toward low-energy regions suggested by continuous relaxations.

Comparing feedback rules under the same mixer clarifies the role of the control law in shaping convergence dynamics. TR-FALQON and SO-FALQON can accelerate descent and improve stability in regimes where first-order updates are less effective, but their practical advantage is most meaningful when paired with drivers that already respect the combinatorial structure of the constraint. This motivates a NISQ-oriented criterion emphasized in our larger-instance test: optimizing hyperparameters to exceed $P_k>0.5$ with the smallest possible number of layers, so that the optimal bitstring becomes unambiguous with a modest sampling budget under realistic depth limitations.

More broadly, the study indicates a promising path for near-term quantum optimization in economically relevant tasks. Beyond the three drivers considered here, exploring additional constraint-aware and hardware-tailored mixers is a natural next step; however, even for fully quantum variational schemes without an outer-loop classical optimization, our results suggest that mixers such as \textcolor{black}{$\hat H_{xy}$} and initializations based on warm-starting already provide a concrete advantage by embedding feasibility and prior information directly into the dynamics. This combination is therefore a compelling building block for future NISQ applications of quantum computing to portfolio selection and related discrete allocation problems.

\backmatter

\bmhead{Acknowledgements}

P.M.P., R.S.C., L.A.M.R. and G.S.F. acknowledge support from the S{\~a}o Paulo Research Foundation (FAPESP; Grants No.~2023/12110-7, 2026/00378-3, 2025/25024-7 and 2025/19585-6). M.C.O. and F.F.F. acknowledge partial financial support from the National Institute of Science and Technology for Applied Quantum Computing through CNPq (Process No.~408884/2024-0) and from FAPESP through the Center for Research and Innovation on Smart and Quantum Materials (CRISQuaM; Process No.~2024/00998-6). F.F.F. and R.S.C. further acknowledges support from ONR (Project No.~N62909-24-1-2012).

\section*{Declarations}

\textbf{Conflict of interest}

The authors declare no competing interests.

\textbf{Ethics approval and consent to participate}

Not applicable.

\textbf{Consent for publication}

All authors reviewed the manuscript and approved its submission for publication.

\textbf{Data availability}


\textcolor{black}{The data used in this study were synthetically generated and are available on GitHub \cite{FALQONPortfolio}.}

\textbf{Materials availability}

Not applicable.

\textbf{Code availability}

\textcolor{black}{The code used in this study is available on GitHub \cite{FALQONPortfolio}.}

\textbf{Author contribution}

P.M.P., L.A.M.R., and R.S.C. conducted the investigation, implemented the numerical simulations, analyzed the results, and wrote the first draft of the manuscript. L.G.E.A. and G.S.F. contributed to the methodology, the discussion of the results, and the revision of the manuscript. M.C.O. and F.F.F. supervised the project, contributed to the conceptualization and interpretation of the results, and revised the manuscript. All authors read and approved the final version of the manuscript.

\bibliography{sn-bibliography}

\end{document}